\documentclass[prl,twocolumn,superscriptaddress,preprintnumbers,showpacs]{revtex4}
\usepackage{amsmath,amssymb,bm,epsfig}

\renewcommand\d{\partial}
\newcommand\grad{\bm{\nabla}}
\newcommand\p{{\bm{p}}}
\newcommand\q{{\bm{q}}}
\renewcommand\k{{\bm{k}}}
\newcommand\ep{\varepsilon_\p}
\newcommand\eq{\varepsilon_\q}
\newcommand\ek{\varepsilon_\k}
\newcommand\+{\dagger}
\newcommand\<{\langle}
\renewcommand\>{\rangle}
\newcommand\eF{\varepsilon_{\mathrm{F}}}

\newcommand{\sect}[1]{\emph{#1.}---}

\begin{document}
\preprint{INT-PUB 06-08, TKYNT-06-06}

\title{$\epsilon$ expansion for a Fermi gas 
at infinite scattering length}
\author{Yusuke~Nishida}
%\email{nishida@nt.phys.s.u-tokyo.ac.jp}
%\homepage{http://tkynt2.phys.s.u-tokyo.ac.jp/~nishida/}
\affiliation{Department of Physics, University of Tokyo,
             Tokyo 113-0033, Japan}
\affiliation{Institute for Nuclear Theory, University of Washington,
             Seattle, Washington 98195-1550, USA}
\author{Dam~Thanh~Son}
\affiliation{Institute for Nuclear Theory, University of Washington,
             Seattle, Washington 98195-1550, USA}

\begin{abstract}
We show that there exists a systematic expansion around four spatial
dimensions for Fermi gas in the unitarity regime. We perform the
calculations to leading and next-to-leading orders in the expansion
over $\epsilon=4-d$, where $d$ is the dimensionality of space. We
find the ratio of chemical potential and Fermi energy to be
$\mu/\eF=\frac12\epsilon^{3/2}+\frac1{16}\epsilon^{5/2}\ln\epsilon
-0.0246\,\epsilon^{5/2}+\cdots$ and
the ratio of the gap in the fermion quasiparticle spectrum and the
chemical potential to be $\Delta/\mu=2\epsilon^{-1}-0.691+\cdots$.
The minimum of the fermion dispersion curve is located at
$|\p|=(2m\varepsilon_0)^{1/2}$ where $\varepsilon_0/\mu=2+O(\epsilon)$. 
Extrapolation to $d=3$ gives results consistent with Monte Carlo
simulations. 
\end{abstract}

\date{April 2006}

\maketitle

\sect{Introduction}
Dilute Fermi gas at infinite scattering length~\cite{Leggett,Nozieres}
has attracted considerable attention recently.  The system can be
realized in atomic traps using the Fesh\-bach
resonance~\cite{OHara,Jin,Grimm,Ketterle,Thomas,Salomon,Thomas05}.
%\cite{experiments}.
It might be relevant for the physics of
dilute neutron gas~\cite{Bertsch}.  It has been suggested that its
understanding may be important for the understanding of high-$T_c$
superconductivity~\cite{highTc}.  From the theoretical perspective
this is a unique nonrelativistic system that has no intrinsic scale
parameter.

Theoretical treatment of the system is difficult, however, precisely
due to the lack of the any small dimensionless quantity.  The usual
Green's function techniques of many-body physics become completely
unreliable since the expansion parameter is large, $na^3\gg1$.  So
far, no systematic treatment has emerged.  Recently considerable
progress has been made by Monte Carlo
simulations~\cite{Carlson2003,Chen:2003vy,Astrakharchik2004,Carlson:2005kg}.
However, there are many reasons that make an analytical treatment, if
it exists, extremely useful.  First, there are many problems that
still cannot be solved by Monte Carlo simulations.  Examples include
polarized Fermi gases, recently realized in
experiments~\cite{Ketterle-polarized,Hulet-polarized}, but whose
lattice realization suffers a fermion sign problem, and questions
related to dynamics like the dynamical response function and the
kinetic coefficients.  Second, in many cases analytical approaches
give unique insights that are not obvious from numerics.

In this Letter we propose an approach based on an expansion around four
spatial dimensions.  In this approach, one would be doing calculation
in $4-\epsilon$ spatial dimensions, where the small number $\epsilon$
is used as a parameter of the perturbative expansion.  Results for the
physical case of three spatial dimensions are obtained by
extrapolating the series expansions to $\epsilon=1$.  This approach
has been extremely fruitful in the theory of the second order phase
transition~\cite{WilsonKogut}.  In our case, we find that even at
$\epsilon=1$ the series over $\epsilon$ is reasonably well-behaved,
strongly suggesting that the limit $d\to4$ is not only theoretically
interesting but also practically useful.

The special role of four spatial dimensions has been recognized by
Nussinov and Nussinov~\cite{nussinov04}.  They noticed that at
infinite scattering length, the two-body wavefunction has a $1/r^2$
behavior when the separation between two fermions $r$ becomes small.
The normalization integral of the wavefunction has a logarithmic
singularity at $r\to0$, from which it is concluded that at $d=4$ the
system must become a noninteracting Bose gas.  
%In particular, at fixed mean particle separation the ground state
%energy should approach zero as the dimension approaches 4. 
As far as we know, no other attempt to exploit this special property
of four dimensions has been made prior to our work.

\sect{Feynman rules and the counting of the powers of $\epsilon$}
Due to the universality of the unitary Fermi gas, any short-range
two-body interaction can be used, if it corresponds to the infinite
scattering length. In particular, we can choose to work with the
Lagrangian of local four-Fermi interaction. 
After a Hubbard-Stratonovich transformation, the Lagrangian density of
the unitary Fermi gas can be written as (here and below $\hbar=1$)
\begin{align}
\begin{split}
 \mathcal{L} &= \Psi^\+\left(i\d_t + \frac{\sigma_3\grad^2}{2m}\right)\Psi
 + \mu\Psi^\dagger\sigma_3\Psi \\ &\qquad\qquad\quad
 - \frac1{c_0} \phi^*\phi
 + \Psi^\+\sigma_+\Psi\phi + \Psi^\+\sigma_-\Psi\phi^*,
\end{split}
\end{align}
where $c_0$ is chosen to correspond to infinite scattering length.  In
dimensional regularization, which we will use, $c_0=\infty$.  From now
on we set $1/c_0=0$.  Here $\Psi$ is a two-component Nambu--Gor'kov
field, $\Psi=(\psi_\uparrow,\psi^\+_\downarrow)^T$,
$\sigma_\pm=\frac12(\sigma_1\pm i\sigma_2)$, and $\sigma_{1,2,3}$ are
the Pauli matrices.

\begin{figure}[tp]
 \includegraphics[width=0.45\textwidth,clip]{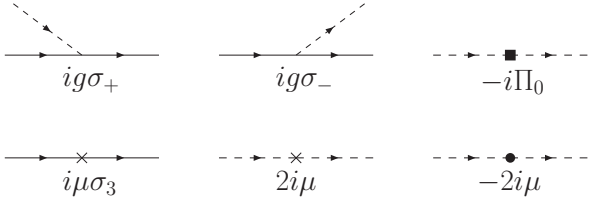}
 \caption{Feynman rules. The two vertices on the last column come
  from $\mathcal{L}_2$, while the rest from $\mathcal{L}_1$. Solid
 (dotted) lines represent the fermion (boson) propagator $iG$ $(iD)$. 
 \label{fig:feynman_rules}}  
%\vspace{-1ex}
\end{figure}

The ground state is a superfluid state where $\phi$ condenses:
$\<\phi\>=\phi_0$. We choose $\phi_0$ to be real. Then we expand 
\begin{equation}
  \phi=\phi_0+ g\varphi, \qquad g = \frac{(8\pi^2\epsilon)^{1/2}}m
  \left(\frac{m\phi_0}{2\pi}\right)^{\epsilon/4},
\end{equation}
where $g\sim O(\epsilon^{1/2})$ was chosen for later convenience,
and rewrite the Lagrangian density as a sum: 
$\mathcal{L}=\mathcal{L}_0+\mathcal{L}_1+\mathcal{L}_2$, where 
\begin{align}
 \begin{split}
  \mathcal{L}_0 & = \Psi^\+\left(i\d_t + \frac{\sigma_3\grad^2}{2m}
  + \sigma_+\phi_0 + \sigma_-\phi_0\right)\Psi \\ %^*
  & \qquad + \varphi^*\left(i\d_t+\frac{\grad^2}{4m}\right)\varphi\,, 
 \end{split} \\
 \mathcal{L}_1 & = g\Psi^\+\sigma_+\Psi\varphi 
 + g\Psi^\+\sigma_-\Psi\varphi^* 
 +\mu\Psi^\+\sigma_3\Psi + 2\mu\varphi^*\varphi\,,
 \phantom{\int}\hspace{-4.5mm}\\
 \mathcal{L}_2 & = -\varphi^*\left(i\d_t
 +\frac{\grad^2}{4m}\right)\varphi - 2\mu\varphi^*\varphi\,.
\end{align}
The part $\mathcal{L}_0$ is a Lagrangian density of noninteracting
fermion quasiparticles and a boson with mass $2m$, whose kinetic terms
are introduced by hand in $\mathcal{L}_0$ and taken out in
$\mathcal{L}_2$. The propagators are generated by $\mathcal{L}_0$ and
the vertices by $\mathcal{L}_1$ and $\mathcal{L}_2$. The fermion
propagator is a $2\times2$ matrix, 
\begin{equation}
  G(p_0,\p) = \frac1{p_0^2-E_\p^2+i\delta}\left(
   \begin{array}{cc}
      p_0 + \ep & -\phi_0 \\
      -\phi_0 & p_0-\ep
   \end{array}
  \right),
\end{equation}
where $\ep=p^2/2m$ and $E_\p=(\ep^2+\phi_0^2)^{1/2}$.
The boson propagator is
\begin{equation}
  D(p_0,\p) = \left(p_0 
  - \frac{\ep}2 + i\delta\right)^{-1}.
\end{equation}
The vertices come from $\mathcal{L}_1$ and $\mathcal{L}_2$ and are
depicted in Fig.~\ref{fig:feynman_rules}, where
\begin{equation}
  \Pi_0 = p_0 -\frac{\ep}2.
\end{equation}
%The two vertices in the last column in Fig.~\ref{fig:feynman_rules}
%come from $\mathcal{L}_2$.  
The fermion-boson coupling is proportional to $g$
and is small in the limit $\epsilon\to0$.

\begin{figure}[bp]
 \includegraphics[width=0.45\textwidth,clip]{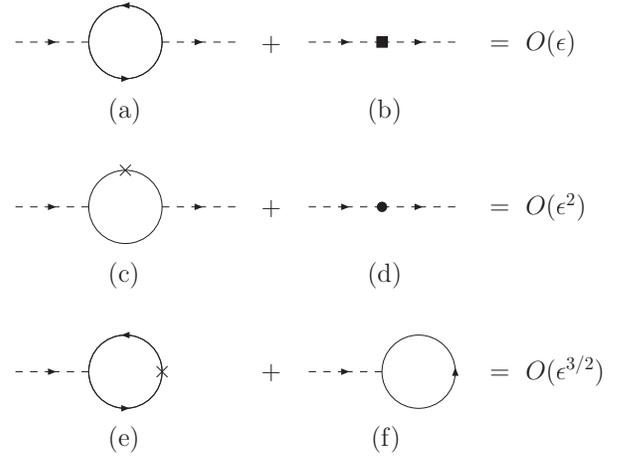}
 \caption{Restoration of naive $\epsilon$ counting for the boson
 self-energy and the cancellation of tadpole diagrams. 
 The fermion loop in (c) goes around clockwise and counterclockwise.
 \label{fig:cancel}}
%\vspace{-1ex}
\end{figure}

\begin{figure}[tp]
 \includegraphics[width=0.40\textwidth,clip]{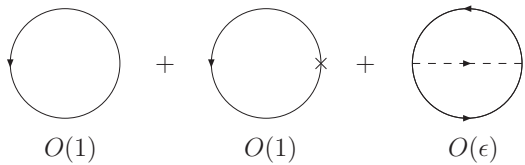}
 \caption[]{Vacuum diagrams for the effective potential up to 
 next-to-leading order in $\epsilon$. The second diagram is $O(1)$
 instead of naive $O(\epsilon)$ because of the $1/\epsilon$
 singularity. \label{fig:potential}}
%\vspace{-1ex}
\end{figure}

Let us first consider Feynman diagrams constructed from $\mathcal{L}_0$
and $\mathcal{L}_1$ only, without the vertices from $\mathcal{L}_2$.  We
make a prior assumption $\mu/\phi_0\sim\epsilon$, which will be checked,
and consider $\phi_0$ to be $O(1)$.  Each pair of boson-fermion vertices 
brings a factor of $\epsilon$, as each $\mu$ insertion.  Therefore the
naive power of $\epsilon$ for a given diagram is $N_g/2+N_\mu$, where
$N_g$ is the number of vertices and $N_\mu$ is the number of $\mu$
insertions. However, this naive counting does not take into account the
fact that there might be inverse powers of $\epsilon$ coming from 
integrals which diverge at $d=4$.  Using a power counting similar to
that in relativistic field theories, one can show that inverse powers of
$\epsilon$ appear only in diagrams with no more than three external
legs. Moreover, from the analytic properties of $G(p)$ and $D(p)$ in the
ultraviolet region, one can show that there are only four diagrams which
have $1/\epsilon$ singularity near four dimensions. They are one-loop
diagrams of the boson self-energy [Figs.~\ref{fig:cancel}(a) and
\ref{fig:cancel}(c)], $\varphi$ tadpole [Fig.~\ref{fig:cancel}(e)], and 
vacuum (the middle of Fig.~\ref{fig:potential}). The diagrams in
Figs.~\ref{fig:cancel}(a) and \ref{fig:cancel}(c) combine with the
vertices from $\mathcal{L}_2$ to restore the naive $\epsilon$ power
counting. 

For example, the diagram in Fig.~\ref{fig:cancel}(a) is
\begin{equation}
%\begin{split}
 -i\Pi(p) 
%  & = -g^2\int\!\frac{dk}{(2\pi)^{d+1}}\, 
%  \Tr\left[G\left(k-\frac p2\right)
%      \sigma_+G\left(k+\frac p2\right)\sigma_-\right].\\
  = -g^2\int\!\frac{dk}{(2\pi)^{d+1}}\, 
  G_{11}\!\left(k-\frac p2\right)G_{22}\!\left(k+\frac p2\right).\\ 
% & = i\int\!\frac{d\k}{(2\pi)^d}\,
%  \left( 2\ek - p_0 + \frac{\ep}2 \right)^{-1}
%\end{split}
\end{equation}
The integral has a pole at $d=4$, so it is $O(1)$ instead of
$O(\epsilon)$ according to the naive counting. 
The residue at the pole can be computed as
\begin{equation}
 \Pi(p) = - %g^2\frac{m^2}{8\pi^2\epsilon}
 \left(p_0-\frac{\ep}2\right) + O(\epsilon),
\end{equation}
which is canceled out exactly by the vertex $\Pi_0$ in $\mathcal{L}_2$. 
Therefore the sum of Figs.~\ref{fig:cancel}(a) and \ref{fig:cancel}(b)
is $O(\epsilon)$. 

Similarly, the diagram in Fig.~\ref{fig:cancel}(c) contains a
$1/\epsilon$ singularity, and is $O(\epsilon)$ instead of naive 
$O(\epsilon^2)$.  The leading part of this diagram is canceled
out by the second vertex from $\mathcal{L}_2$, and the total is again 
$O(\epsilon^2)$. 

Finally, the $\varphi$ tadpole diagram with one $\mu$ insertion
[Fig.~\ref{fig:cancel}(e)] is $O(\epsilon^{1/2})$ instead of naive
$O(\epsilon^{3/2})$.  The only diagram that can cancel this is the
tadpole diagram with no $\mu$ insertion, Fig.~\ref{fig:cancel}(f).  The 
condition of cancellation determines $\phi_0(\mu)$ to leading order in
$\epsilon$.  This condition will be automatically satisfied by the
minimization of the effective potential.

Thus, we can now develop a diagrammatic technique for our system.  For
any Green's function, we write down all Feynman diagrams according to
the Feynman rules, using the propagators from $\mathcal{L}_0$ and the
vertices from $\mathcal{L}_1$.  If there is any subdiagram of the type
in Fig.~\ref{fig:cancel}(a) and Fig.~\ref{fig:cancel}(c), we add a
diagram with a vertex from $\mathcal{L}_2$.  The result will be 
$O(\epsilon^{N_g/2+N_\mu})$~\footnote{At a sufficiently high order in
the perturbation theory ($\epsilon^3$ compared to the leading order), 
a resummation of the boson propagator is needed to avoid 
infrared singularities.}.

%\sect{Leading order results}
%The gap equation is obtained from the cancellation of tadpole diagrams.
%To leading $O(1)$ order there are two diagrams,
%Figs.~\ref{fig:cancel}(e)-(f), and the solution to the gap equation is
%\begin{equation}
%  \frac{\phi_0}\mu = \frac2\epsilon + O(1).
%\end{equation}
%The ratio of the chemical potential and the Fermi energy is
%\begin{equation}
%  \xi = \frac12\epsilon^{3/2}.
%\end{equation}
%To leading order, the results coincide with those obtained from the
%BCS mean field theory.

\sect{Leading and next-to-leading order results}
We shall perform explicit calculations, employing the Feynman rules
and the $\epsilon$ power counting that have just been developed, to
leading and next-to-leading orders.  The dependence of $\phi_0$ on
$\mu$ is most conveniently computed from the minimization of the
effective potential $V_\mathrm{eff}(\phi_0)$~\cite{Peskin:1995ev}.  To 
next-to-leading order, the effective potential receives contribution
from three vacuum diagrams drawn in Fig.~\ref{fig:potential}: fermion
loops with and without a $\mu$ insertion and a fermion loop with the
boson exchange.  The contribution from the one-loop diagrams reads
\begin{align}
\begin{split}
 V_1(\phi_0) &=
 \frac{\phi_0}3\left[1+\frac{7-3(\gamma+\ln2)}6\epsilon\right] 
 \left(\frac{m\phi_0}{2\pi}\right)^{d/2}\\
 &\ \ -\frac\mu\epsilon\left[1+\frac{1-2(\gamma-\ln2)}4\epsilon\right]
 \left(\frac{m\phi_0}{2\pi}\right)^{d/2},
\end{split}
\end{align}
where $\gamma\approx0.57722$ is the Euler-Mascheroni constant.
The contribution of the two-loop diagram is
\begin{align}
% \begin{split}
  V_2(\phi_0) 
%  & = g^2\int\!\frac{dp\,dq}{(2\pi)^{2d+2}}
%  \Tr\left[G(p)\sigma_+G(q)\sigma_-\right]D(p-q)\\
  & = g^2\int\!\frac{dp\,dq}{(2\pi)^{2d+2}}\,
  G_{11}(p)G_{22}(q)D(p-q)\\
  & = -\frac{g^2}4\! \int\!\frac{d\p\,d\q}{(2\pi)^{2d}}\,
  \frac{(E_\p-\ep)(E_\q-\eq)}
  {E_\p E_\q (E_\p+E_\q+\varepsilon_{\p-\q}/2)}. \notag
% \end{split}
\end{align}
This integral is convergent even at $d=4$.  Its value is
\begin{equation}
  V_2(\phi_0)=-C\epsilon\left(\frac{m\phi_0}{2\pi}\right)^{d/2}\phi_0,
\end{equation}
where the constant $C$ is given by a two-dimensional integral
\begin{align}
\begin{split}
  C &= %\int\limits_0^\infty\!dx\!\int\limits_0^\infty\!dy\, 
 \int_0^\infty\!dx\int_0^\infty\!dy\, 
 \frac{[f(x)-x][f(y)-y]}{f(x)f(y)} \\ &\qquad\qquad\qquad 
 \times\left[g(x,y) - \sqrt{g^2(x,y)-xy}\right]
\end{split}
\end{align}
with $f(x)=(x^2+1)^{1/2}$ and $g(x,y)=f(x)+f(y)+\frac12(x+y)$.  
The result of the numerical integration is
\begin{equation}
  C\approx 0.14424.  
\end{equation}

The minimum of the effective potential
$V_\mathrm{eff}(\phi_0)=V_1(\phi_0)+V_2(\phi_0)$ is located at 
\begin{equation}\label{phi0mu}
  \phi_0 = \frac{2\mu}\epsilon [1 + (3C-1+\ln 2)\epsilon].
\end{equation}
Note that the previously made assumption $\mu/\phi_0=O(\epsilon)$ is
now checked.  Also if one used the mean field approximation, one would 
reproduce the leading $2\mu/\epsilon$ term in Eq.~(\ref{phi0mu}),
but not the $O(\epsilon)$ correction.  The value of $V_\mathrm{eff}$ at 
$\phi_0$ in Eq.~(\ref{phi0mu}) determines the pressure 
$P=-V_\mathrm{eff}(\phi_0)$ at chemical potential $\mu$. 
%\begin{equation}\label{Pmu}
% P =\frac{\phi_0}6 \left(\frac{m\phi_0}{2\pi}\right)^{d/2}
%  \left[ 1+ \left(\frac{17}{12}-3C-\frac{\gamma+\ln 2}2\right)
%   \epsilon\right],
%\end{equation}
%where the dependence of $\phi_0$ on $\mu$, Eq.~(\ref{phi0mu}), has
%been substituted.  
The density is determined from $n=\d P/\d\mu$, 
%\begin{equation}
%  n = \frac{\d P}{\d\phi_0}\frac{\d\phi_0}{\d\mu} = 
%  \frac1\epsilon\left(\frac{m\phi_0}{2\pi}\right)^{d/2}
%  \left[1+\frac{2-2\gamma+2\ln2}4\epsilon\right]
%\end{equation}
and the Fermi energy from the thermodynamic of free gas in $d$
dimensions is given by
\begin{equation}
  \eF = \frac{2\pi}m \left[ \frac12\Gamma\left(\frac d2+1\right)n
  \right]^{2/d} = \frac{\phi_0}{\epsilon^{2/d}}
  \left(1-\frac{1-\ln2}4\epsilon\right).
\end{equation}
The nontrivial power of $\epsilon$ comes from taking
$n\sim\epsilon^{-1}$ to the power of $2/d$.  We find the parameter
$\xi\equiv\mu/\eF$,
\begin{equation}
  \xi = \frac{\epsilon^{3/2}}2 
  \exp\!\left(\frac{\epsilon\ln\epsilon}{8-2\epsilon}\right)
  \left[ 1- \left(3C -\frac54 (1-\ln2)\right)\epsilon\right].
\end{equation}
Substituting the numerical value for $C$, one finds
\begin{equation}
  \xi = \frac12\epsilon^{3/2} + \frac1{16}\epsilon^{5/2}\ln\epsilon
  - 0.0246\,\epsilon^{5/2} + \cdots. %0.0245793
\end{equation}
The smallness of the coefficient in front of $\epsilon^{5/2}$ is a
result of a cancellation between the two-loop correction and the
subleading terms from the expansion of the one-loop diagrams around 
$d=4$.

\begin{figure}[tp]
 \includegraphics[width=0.45\textwidth,clip]{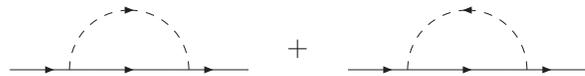}
 \caption{One-loop diagrams for the fermion self-energy
 of order $O(\epsilon)$. \label{fig:self_energy}}
%\vspace{-1ex}
\end{figure}

\sect{Quasiparticle spectrum}
To leading order, the dispersion relation
of fermion quasiparticles is
$\omega(\p)=E_\p=(\ep^2+\phi_0^2)^{1/2}$.  It has a minimum at
$\p=\bm0$ with a gap equal to $\Delta=\phi_0=2\mu/\epsilon$. 
The next-to-leading order correction comes from three sources: from the 
correction of $\phi_0$ in Eq.~(\ref{phi0mu}), from a $\mu$ insertion to
the fermion propagator, and from the one-loop self-energy diagrams,
$-i\Sigma(p)$, depicted in Fig.~\ref{fig:self_energy}.  Using the
Feynman rules one can see that there are corrections only to the
diagonal elements of the self energy: 
\begin{equation}
\begin{split}
  \Sigma_{11}(p) &= ig^2\!\int\!\frac{dk}{(2\pi)^{d+1}}\,
  G_{22}(k)D(p-k)\\
  &=-\frac{g^2}2\!\int\!\frac{d\k}{(2\pi)^d}\, \frac{E_\k-\ek}
  {E_\k (E_\k + \varepsilon_{\k-\p}/2-p_0)}
\end{split}
\end{equation}
and $\Sigma_{22}(p_0,\p)=-\Sigma_{11}(-p_0,\p)$. 
To find the correction to the dispersion relation around its minimum, we
only have to evaluate the self-energy at $p_0=\phi_0$, $\ep\sim\mu$ and 
expand $\Sigma(p_0,\p)=\Sigma^0(\phi_0,\bm0)
+\Sigma'(\phi_0,\bm0)\,\ep/\phi_0$.
%and performing the $\k$ integration, we have
%\begin{multline}
% \Sigma_{11}(\phi_0,\p)=\epsilon\left(2-8\ln3+8\ln2\right)\phi_0\\
% +\epsilon\left(-\frac83+8\ln3-8\ln2\right)\ep,
%\end{multline}
%\begin{multline}
% \Sigma_{22}(\phi_0,\p)=\epsilon\left(-2-8\ln3+16\ln2\right)\phi_0\\
% +\epsilon\left(-\frac73-8\ln3+16\ln2\right)\ep.
%\end{multline}
By solving the equation 
$\det[G^{-1}(\omega,\p)+\mu\sigma_3-\Sigma(\omega,\p)]=0$ in terms of
$\omega$, we see the dispersion relation around its minimum is given by 
\begin{equation}
 \omega(\p)=\Delta+\frac{(\ep-\varepsilon_0)^2}{2\phi_0},
\end{equation}where 
$\Delta=\phi_0+(\Sigma_{11}^0+\Sigma_{22}^0)/2$ and
$\varepsilon_0=\mu+(\Sigma_{22}^0-\Sigma_{11}^0)/2
-(\Sigma_{11}'+\Sigma_{22}')/2$.
%\begin{equation}
% \Delta = \phi_0 + \frac{\Sigma_{11}^0+\Sigma_{22}^0}2 
%\end{equation}
%\begin{equation}
% \varepsilon_0 = \mu + \frac{\Sigma_{22}^0-\Sigma_{11}^0}2 
%  - \frac{\Sigma_{11}'+\Sigma_{22}'}2.
%\end{equation}

The result of an explicit calculation is the following.  
The minimum of the dispersion curve is located at a nonzero value of
momentum, $|\p|=(2m\varepsilon_0)^{1/2}$, where
\begin{equation}
 \varepsilon_0=2\mu. 
  %\varepsilon_0\equiv\frac{p_0^2}{2m}=2\mu.
\end{equation}
Note the difference with the mean field approximation, in which
$\varepsilon_0=\mu$. The correction to the gap is
\begin{equation}
  \frac12[\Sigma_{11}(\phi_0,\bm0)+\Sigma_{22}(\phi_0,\bm0)]
  = -\epsilon\left(8\ln3-12\ln2\right)\phi_0.
\end{equation}
Combining it with the correction in Eq.~(\ref{phi0mu}), we obtain
\begin{equation}
  \frac\Delta\mu 
   = \frac2\epsilon\left[1+(3C-1-8\ln3+13\ln2)\epsilon\right]
  \approx \frac2\epsilon-0.691. %0.690521
\end{equation}

\sect{Extrapolation to $\epsilon=1$}
Although the formalism is based on the smallness of $\epsilon$, we see
that even at $\epsilon=1$ the corrections are reasonably small.  If we
naively use only the leading and next-to-leading order results,
extrapolation to $\epsilon=1$ gives for three spatial dimensions
\begin{equation}
 \xi \approx 0.475, \qquad \frac{\varepsilon_0}{\mu} \approx 2, 
  \qquad \frac{\Delta}{\mu}\approx 1.31. %0.475421 and 1.30948
\end{equation}
They are reasonably close to the results of recent Monte Carlo 
simulations, which yield $\xi\approx0.42$, $\varepsilon_0/\mu\approx1.9$,
and $\Delta/\mu\approx1.2$~\cite{Carlson:2005kg}.  
They are also consistent with recent experimental measurements of $\xi$,
where $\xi=0.51\pm0.04$~\cite{Thomas05} and
$\xi=0.46\pm0.05$~\cite{Hulet-polarized}. 
Thus there is a strong indication that the $\epsilon$ expansion is
useful in practice. A calculation of the $\epsilon^2$ corrections to
these results would be extremely interesting.

\sect{Conclusion}
We have developed a systematic expansion, treating the dimensionality of
space as close to four, and obtained very reasonable results. As far as
we know, this is the only systematic expansion for the unitary Fermi gas
at zero temperature that exists at this moment.  We found that the the 
corrections are not too big even when extrapolated to $\epsilon=1$, 
which suggests that the picture of the unitary Fermi gas as a collection
of weakly interacting fermionic and bosonic quasiparticles may be a
useful starting point even in three spatial dimensions.  There is a host
of problems that can be addressed using this approach: the phase diagram 
of the polarized system, the structure of the superfluid vortex,
finite-temperature physics, etc. It is interesting to note that the
critical dimension of a superfluid-normal phase transition is also four,
making weak-coupling calculations reliable at any temperature for small
$\epsilon$.

\sect{Acknowledgment}
The authors thank P.~Arnold, M.~M.~Forbes, G.~Rupak, M.~A.~Stephanov,
and A.~Vuorinen for discussions. Y.~N. is supported by the Japan Society
for the Promotion of Science for Young Scientists.  This work is
supported, in part, by DOE Grant No.\ DE-FG02-00ER41132.

\end{document}